# $Fe^{3+}$-$\eta^2$-peroxo species in Superoxide Reductase from *Treponema pallidum*. Comparison with *Desulfoarculus baarsii* †

Christelle Mathé[‡§], Vincent Nivière[§*], Chantal Houée-Levin[⊥], and Tony A. Mattioli[‡*]

† TAM gratefully acknowledges an equipment grant from the Region Council of the Ile-de France (S.E.S.A.M.E.)

‡ Laboratoire de Biophysique du Stress Oxydant, SBE and CNRS URA 2096, Département de Biologie Joliot Curie, CEA Saclay, 91191 Gif-sur-Yvette cedex, France

§ Laboratoire de Chimie et Biochimie des Centres Redox Biologiques, DRDC-CEA/CNRS/Université Joseph Fourier, 17 avenue des Martyrs, 38054 Grenoble cedex 9, France

⊥ Laboratoire de Chimie Physique, CNRS/Université Paris-Sud, Bâtiment 350, Centre Universitaire 91405 Orsay Cedex, France

* To whom correspondence should be addressed:

Tony A. Mattioli

Tel.: +33-1-69084166; FAX: +33-1-69088717; E-mail: tony.mattioli@cea.fr

Vincent Nivière

Tel.: +33-4-38789109; FAX: +33-4-38789124; E-mail: vniviere@cea.fr

**Running Title**: $Fe^{3+}$-peroxo species in *Treponema pallidum* SOR.

**Footnotes**

[1]Abbreviations: SOR, superoxide reductase; EPR, electron paramagnetic resonance; WT, wild type; RR, resonance Raman.

**Synopsis**

Superoxide reductases (SORs) are superoxide ($O_2^{\bullet-}$)-detoxifying enzymes that catalyse the reduction of $O_2^{\bullet-}$ into hydrogen peroxide. Three different classes of SOR have been reported on the basis of the presence or not of an additional N-terminal domain. They all share a similar active site, with an unusual non-heme Fe atom coordinated by four equatorial histidines and one axial cysteine residues. Crucial catalytic reaction intermediates of SOR are purported to be $Fe^{3+}$-(hydro)peroxo species. Using resonance Raman spectroscopy, we compared the vibrational properties of the $Fe^{3+}$ active site of two different classes of SOR, from *Desulfoarculus baarsii* and *Treponema pallidum*, along with their ferrocyanide and their peroxo complexes. In both species, rapid treatment with $H_2O_2$ results in the stabilization of a side-on high spin $Fe^{3+}$-($\eta^2$-OO) peroxo species. Comparison of these two peroxo species reveals significant differences in vibrational frequencies and bond strengths of the Fe-$O_2$ (weaker) and O-O (stronger) bonds for the *T. pallidum* enzyme. Thus, the two peroxo adducts in these two SORs have different stabilities which are also seen to be correlated with differences in the Fe-S coordination strengths as gauged by the Fe-S vibrational frequencies. This was interpreted from structural variations in the two active sites, resulting in differences in the electron donating properties of the *trans* cysteine ligand. Our–results suggest that the structural differences observed in the active site of different classes of SORs should be a determining factor for the rate of release of the iron-peroxo intermediate during enzymatic turnover.

## INTRODUCTION

Superoxide reductase (SOR[1]), a small metalloprotein of ca. 14 kDa, is a newly discovered antioxidant enzyme which catalyses the one-electron reduction of superoxide anion ($O_2^{\bullet -}$) to form hydrogen peroxide $H_2O_2$ [1-3]:

$$O_2^{\bullet -} + 1e^- + 2H^+ \rightarrow H_2O_2$$

SOR proteins have been found in several air-sensitive bacteria [2,4,5] and archaea [1,6] and their activity is thought to play a fundamental role for these anaerobic organisms in the defence against oxidative and superoxide stress or exposure to adventitious molecular oxygen [7,8].

The active site of SOR consists of a non-heme $Fe^{2+}$ centre (named Centre II in those 2Fe-containing SORs) in an unusual [$His_4$ $Cys_1$] square pyramidal pentacoordination [2,9,10]. In the reduced state, the sixth, axial coordinating site of the iron is vacant, and suggests the most obvious site for initial binding of the superoxide substrate [9,12]. SOR reacts specifically at a nearly diffusion-controlled rate with $O_2^{\bullet -}$, generating $H_2O_2$ and the oxidized $Fe^{3+}$ active site [2]. In the oxidized state, the active site $Fe^{3+}$ centre becomes hexacoordinated, with its sixth coordination site occupied by a conserved glutamic acid residue (E47 in *Desulfoarculus baarsii*) [12-14]. The electrons involved in the reaction can be provided to the oxidized active site by rubredoxin [1,15] or by NAD(P)H dependent cellular reductases in the absence of rubredoxin [2].

Three classes of SORs have been described so far, all of which display a similar active site. *Desulfovibrio desulfuricans* [16], *Desulfovibrio vulgaris* Hildenborough [17] and *Desulfoarculus baarsii* [2] SORs are representatives of Class I, in which the catalytic domain is linked to a small N-terminal domain structurally similar to desulforedoxin and containing an additional iron centre, named Centre I [9]. The mononuclear $Fe^{3+}$ of Centre I is ligated with four cysteines in a distorted tetrahedral arrangement, of rubredoxin-type. SORs of Class II, such as *Pyrococcus furiosus* SOR [1], are characterized by the absence of the additional

N-terminal domain [12]. In Class III as represented by the *Treponema pallidum* SOR [4,5], this domain is present but is incapable of binding the second Fe atom and therefore Centre I is absent. In fact, when present, a functional role of Centre I has not been established [2,4,10,11].

Pulse radiolysis studies on SORs have shed light on their enzyme mechanism [19-24]. It is now generally agreed that the reaction of $O_2^{\bullet-}$ with the reduced iron Centre II proceeds through an inner-sphere mechanism. The first step of the reaction consists of a bi-molecular reaction of SOR with $O_2^{\bullet-}$ in a nearly diffusion-controlled process ($10^9$ $M^{-1}$ $s^{-1}$), to form a first reaction intermediate. The subsequent step was clearly identified as a protonation process [23,24], leading to the formation of a second reaction intermediate [24], before the formation of the final products of the reaction, $H_2O_2$ and the oxidized iron Centre II. However, this last step of the reaction mechanism is not kinetically characterized yet [20,21,24]. Other groups propose that the first intermediate leads directly to the final product of the reaction, without formation of a second intermediate [19,22,23]. In all cases, the last step of the reaction might proceed with the intervention of the conserved glutamate residue, which becomes the sixth ligand of the $Fe^{3+}$ ion in the oxidized Centre II [12,14,23,24].

The exact chemical nature of these intermediates observed in pulse radiolysis, which may be formulated as $Fe^{2+}$-superoxo, $Fe^{3+}$-peroxo, or $Fe^{3+}$-hydroperoxo, is yet to be firmly established. We have recently identified by resonance Raman spectroscopy a transient high-spin "side-on" $Fe^{3+}$-peroxo species ($Fe^{3+}$-($\eta^2$-OO)) which could be trapped in the SOR active site of *Desulfoarculus baarsii* both in the WT and in the E47A mutant where the conserved Glu 47 residue has been replaced with alanine [25]. This was possible by rapidly mixing and freezing a solution of resting ferrous SOR with a slight excess of $H_2O_2$. These data showed that this SOR active site can readily accommodate a $Fe^{3+}$-peroxo species with a "side-on" configuration. In addition, this $Fe^{3+}$-($\eta^2$-OO) species was sizeably stabilized in the mutant

E47A [25]. This is in accord with the postulated role for the E47 residue in the release of the reaction intermediate from the SOR active site [25].

In this work, in order to determine if different classes of SORs can also accommodate a high-spin $Fe^{3+}$-($\eta^2$-OO) species, we have investigated the formation of the peroxo species in the Class III SOR from *Treponema pallidum*, on the WT and E48A mutant (equivalent of the E47A mutant in *D. baarsii*). Comparison of the two peroxo species between the SORs from *T. pallidum* and *D. baarsii* indicate different stabilities of the peroxo adduct, which can be directly related with differences in the iron-ligand coordination strengths as seen by vibrational Raman spectroscopy.

## EXPERIMENTAL

### Biochemical and chemical reagents.

$K_3Fe(CN)_6$, ammonium persulphate were from Sigma. $K_2IrCl_6$ was from Strem Chemical Inc. $H_2{}^{16}O_2$ (30% in water) was from Aldrich. $H_2{}^{18}O_2$ (2% in water with 90% $^{18}O$) was from ICON Stable Isotopes.

### Site-directed mutagenesis, protein expression and purification.

Two primers were designed for PCR- based site-directed mutagenesis to create the *T. pallidum* SOR mutant E48A. Primer 1 (5'GGA TGC AGC GAA GGC AAA GCA TAT CCC CG3') and primer 2 (5'CGG GGA TAT GCT TTG CCT TCG CTG CAT CC3') contained the mutation of interest (underlined). Mutagenesis was carried out on the plasmid pVN10-2 [4] with the QuickChange$^{TM}$ site-directed mutagenesis kit from Stratagene. The mutation was verified by DNA sequencing. The resulting plasmid, pCMTPE48A was transformed in *E. coli* DH5α. Over-expression and purification of the E48A mutant protein was carried out as reported for the wild-type protein [4]. Purification of the wild-type and E47A proteins from *D. baarsii* were performed as described in [2] and [20]. Purified protein samples, in 10 mM Tris-HCl pH 7.6, were concentrated using Microcon 10 microconcentrators (10 kDa cut-off membranes). Total iron content in proteins was determined using atomic absorption spectroscopy as described elsewhere [2]. Unless otherwise stated, proteins were oxidized using 3 equivalents of $K_2IrCl_6$. For certain cases, 1.5 equivalents of $K_3Fe(CN)_6$ were used. In general, the excess oxidant was removed by washing using Microcon 10 microconcentrators. Ammonium persulphate (1 equivalent) was also used as an oxidant. For the wild-type SOR from *T. pallidum*, complete reduction of the active site was achieved by adding slight excess of sodium ascorbate (Sigma).

In general for the resonance Raman samples, 1 µl of concentrated protein (1-5 mM) in 100 mM of buffer was deposited on glass slide sample holder and then transferred into a cold helium gas circulating optical cryostat (STVP-100, Janis Research) held at 15 K. For the rapidly frozen $H_2O_2$-treated Raman samples, 5 µl of concentrated SOR (either in its $Fe^{2+}$ or $Fe^{3+}$ state) protein (3-6 mM) on ice was manually mixed rapidly with an equal volume of $H_2O_2$ solution (on ice) whose concentration was adjusted in order to obtain a final SOR protein : $H_2O_2$ ratio of *ca.* 1:6 equivalents. 3-5 µL of the resulting mixture was promptly transferred to the glass slide sample holder and immediately immersed in liquid nitrogen before transferring to the optical cryostat for Raman measurement. The entire mixing/freezing operation took no more than ca. 3 seconds. For reaction of $H_2O_2$ with SOR in its ferric $Fe^{3+}$ state, the protein was initially oxidized with 3 eq. of $K_2IrCl_6$, washed with three volumes of buffer using membrane microconcentrators, then concentrated before mixing with 6 eq. of $H_2O_2$. The pH of trial samples (200 µL, 100 mM buffer) were re-measured i) after $H_2O_2$ mixing and ii) after freeze/thawing and were observed to not change within ± 0.1 pH unit at pH 7.5 and 8.5. The resonance Raman spectra of the WT and E47A mutant of *D. baarsii* SOR [26] are essentially indistinguishable when reacted with $H_2O_2$ in their respective $Fe^{2+}$ or $Fe^{3+}$ states (for example, see Supplemental Data, Figure S1) indicating that a similar $Fe^{3+}$-peroxo species is trapped. For the spectra reported in this work, the reaction of $H_2O_2$ with SOR was done in the $Fe^{2+}$ state which circumvents the additional $K_2IrCl_6$-oxidation and subsequent washing steps.

**Pulse radiolysis.**

Pulse radiolysis measurements were performed as described elsewhere [20,21]. Briefly, free radicals were generated by irradiation of $O_2$-saturated aqueous protein solutions (100 µM), in Tris-HCl 10 mM pH 7.6, 0.1 M sodium formate with 200 ns pulses of 4 MeV electrons at the linear accelerator at the Curie Institute, Orsay, France. Superoxide anion, $O_2^{\bullet -}$, was generated during the scavenging by 100 mM formate of the radiolytically produced hydroxyl radical, $HO^{\bullet}$, as previously described [20]. Doses of ca. 5 Gy per pulse resulted in ca. 3 μM of $O_2^{\bullet -}$. Reactions were followed spectrophotometrically, between 450 and 750 nm, at 20°C in a 2 cm path length cuvette. Kinetic traces were analyzed using a Levenberg-Marquardt algorithm from the Kaleidagraph software package (Synergy Software).

**Spectroscopy.**

Optical absorbance measurements were made using a Varian Cary 1 Bio spectrophotometer, in 1 cm path length cuvettes. For measurements using $H_2O_2$ as the oxidant, spectra were rapidly recorded, immediately after addition, using a multichannel diode-array spectrophotometer (Helwett-Packard) and the evolution of the absorption spectrum was followed every few seconds. Low temperature 4.2 K X-band EPR spectra were recorded on a Bruker EMX 081 spectrometer equipped with an Oxford Instrument continuous flow cold He gas cryostat.

Resonance Raman spectra were recorded using a modified single-stage spectrometer (Jobin-Yvon) equipped with a liquid nitrogen cooled back-thinned CCD detector (2000 X 800 pixels), and excitation at 647.1 nm (30 mW) was provided from a Spectra Physics Series 2000 $Kr^+$ laser. A holographic notch filter (Kaiser Optical) was used reject stray light. Spectra were calibrated using the exciting laser line, along with the $SO_4^{2-}$ (983 $cm^{-1}$) and ice (230 $cm^{-1}$) Raman bands from a frozen aqueous sodium sulphate solution. Spectral resolution was < 3 $cm^{-1}$ with entrance slits at 100 μm. Frequency accuracy was ± 1 $cm^{-1}$ and frequency repeatability was 1 pixel (resolution < 2 pixels) for the spectrograph. Reported spectra were the result of the averaging of 40 single spectra recorded with 30 seconds of accumulation time. The reproducibility of the reported Raman frequencies was determined to be ± 1 $cm^{-1}$. Verification of small frequency differences of 2 $cm^{-1}$ of homologous bands originating from two different samples was established by recording, during the same day, the two different samples deposited the same sample holder under the same spectroscopic/geometric conditions. Baseline corrections were performed using GRAMS 32 (Galactic Industries). In all reported spectra, the contributions from ice have been subtracted using the GRAMS 32 software by cancellation of the 230 $cm^{-1}$ band.

## RESULTS

### Characterization of the *T. pallidum* E48A mutant.

The *Treponema pallidum* SOR E48A mutant was purified in an homogeneous form, with an iron content of 0.75 iron atom per polypeptide chain. The E48A mutant was isolated in a completely reduced state, stable in the presence of air. Figure 1A shows the 400-900 nm region of the electronic absorption spectrum of the *T. pallidum* E48A mutant in its ferric state oxidized with $K_2IrCl_6$ at pH 8.5. Similar spectra have been observed when it is oxidized with ammonium persulphate or superoxide generated with the xanthine / xanthine oxidase system (data not shown). As described in the case of the *Desulfoarculus baarsii* enzyme, the maximum of the absorption spectrum depends on the pH. Upon increasing the pH from 5 to 9.1, the absorption band of the iron center exhibits a 80 nm blue shift from 650 to 560 nm, with a $pK_a$ value of 6.0 (Fig. 1B). This value is almost identical to that reported in the case of the E47A mutant of *D. baarsii* (pKa = 6.7) [24]. On the other hand, between pH 5 and 8.5, the spectrum of the oxidized *T. pallidum* wild-type SOR exhibits the same maximal absorption at 650 nm (compare Figure 1a with Supplemental Data Figure S2a). Above pH 8.5, the iron center becomes unstable and no spectra could be collected. However, these pH-dependent absorption data are similar to those of the wild-type SOR from *D. baarsii*, which exhibits the 80 nm blue-shift absorption transition at higher pH ($pK_a$ 9.0) [*24*].

Figure 2a shows the low temperature X-band electron paramagnetic (EPR) spectrum of the *T. pallidum* E48A mutant oxidized with $K_2IrCl_6$ at pH 8.5. The major feature at ca. g = 4.3 indicates the high spin state of the ferric centre, as previously observed for all other SORs to date [2,4,13,16,27]. There is also a small feature observable at g = 4.2 (Inset Fig. 2a). In addition, the EPR spectrum reveals further heterogeneity in the coordination environment of the ferric centre. The resonance at g = 8.8 arises from a rhombic species (E/D ≈ 0.15) while the resonances at g = 7.4 and 5.7 arise from an axial species (E/D ≈ 0.05). Similar structural heterogeneity and EPR resonances have been also observed for the wild-type *T. pallidum* and *P. furiosus* SORs [13,27]. In contrast, no such iron coordination heterogeneity was observed for wild-type *D. baarsii* or its E47A mutant, both of which were rhombic (E/D = 0.33) when oxidized with $K_2IrCl_6$ or $O_2^{\bullet-}$ [2,20,25].

The reactivity of the *T. pallidum* E48A SOR with $O_2^{\bullet-}$ was studied by pulse radiolysis at pH 7.6. The experimental conditions were identical to those used for the wild-type [21] and for the *D. baarsii* SORs [20,24]. Like the wild-type SOR [21], the mutant reacted very rapidly with superoxide, with a nearly diffusion-limited second-order rate constant $k_1$ = $10^9 M^{-1}s^{-1}$ (Table 1), to form a first intermediate species which exhibits a maximum absorbance at *ca*. 600 nm (data not shown), comparable with that found for the wild-type protein [21]. This first intermediate species decays with a rate constant $k_2$ = 2080 $s^{-1}$ (Table 1) to form a second intermediate absorbing maximally at *ca*. 630 nm, similarly to that determined for the wild-type protein [21]. Finally, the last part of the reaction, the transformation of the second intermediate to the final products, occurs after 30-50 ms of the reaction time, which could not be observed with our instrumentation, as previously mentioned [21,24].

**Resonance Raman Spectroscopy of SOR wild-type and mutated forms oxidized with $K_2IrCl_6$**

Figure 3 shows the 15 K RR spectrum of the wild-type SOR from *T. pallidum* oxidized with $K_2IrCl_6$, excited using 647.1 nm radiation, in resonance with the S → $Fe^{3+}$ charge transfer band of the active site [16,28]. The active site modes thus enhanced predominantly arise from the $Fe^{3+}$-S(Cys) moiety [10,25,28].

Unlike the SOR from *D. baarsii*, this resonance Raman spectrum is not complicated by the preresonance Raman contribution of Center I. Johnson and co-workers, using $^{34}S$, $^{15}N$ and $^{54}Fe$ global protein isotopic labelling of SORs from *P. furiosus* [28] and *D. vulgaris* [10] have shown that the low-frequency Fe-S stretching mode is kinematically coupled to several cysteine ligand deformation modes close in energy, resulting in several mixed modes. The low frequency region of the spectrum of the SOR from *T. pallidum* is dominated by bands at 304, 311, 326, and 357 $cm^{-1}$. Based on the assignments made by Johnson and co-workers [10,28], we may assign the 304, 311, and 326 $cm^{-1}$ bands in Figure 3, respectively, to Fe-S stretching and bending modes (contributions from S-$C_\beta$-$C_\alpha$ bending deformations). The 357 $cm^{-1}$ band can be assigned to a cysteine deformation mode containing significant C-N character [28]. The 500-800 $cm^{-1}$ region is expected to contain contributions from overtone and combination bands, except for the 658 $cm^{-1}$ band which was assigned to a fundamental with significant cysteine C-N character [10,28]. The intense 746 $cm^{-1}$ band in Figure 3b can be assigned to a cysteine S-$C_\beta$ stretching mode [28]. The weakly enhanced bands in the 200-250 $cm^{-1}$ region are attributable to $Fe^{3+}$-N stretching modes from the coordinating histidine residues [10,28]. Bands are observed at 215, 232 (the residual feature in this band is due to the subtraction of the ice band at 230 $cm^{-1}$) and 255 $cm^{-1}$, which agree well with those observed in the *P. furiosus* spectrum which showed $^{15}N$ isotopic shifts [28].

For the *D. baarsii* RR spectrum (Figure 3), the preresonance contributions from Center I have been subtracted by cancellation of the 384 $cm^{-1}$ band, arising uniquely from the non-active Center I site. As with the oxidized *T. pallidum* RR spectrum, the difference spectrum is also dominated by contributions from the $Fe^{3+}$-S(Cys) moiety (Figure 3). By analogy to the assignments made for the SORs from *P. furiosus* [28] and *D. vulgaris* [10], the 299/316/323 $cm^{-1}$ cluster of bands in the *D. baarsii* spectrum may be attributed to primarily Fe-S stretching mode and cysteine bending mode contributions. Since the 299 $cm^{-1}$ band is the most intense, it is most likely predominantly $Fe^{3+}$-S stretching in character. Similarly for *T. pallidum*, the relatively intense 304 $cm^{-1}$ band most likely represents the mode with predominant $Fe^{3+}$-S stretching in character in the *T. pallidum* active site. The 357 $cm^{-1}$ band can be assigned to a cysteine deformation mode and involving N atoms [28]. The 742 $cm^{-1}$ bands in the *D. baarsii* spectrum was assigned to the S-$C_β$ stretching mode and is significantly lower than that observed for *T. pallidum* (746 $cm^{-1}$) and *P. furiosus* (748 $cm^{-1}$) [28]. The weakly enhanced $Fe^{3+}$-N stretching mode bands [28] are seen at 216, 234 and 238 $cm^{-1}$. Also, two shoulders at 277 and 289 $cm^{-1}$ are visible, which corresponds to the broader unresolved shoulder at 280 $cm^{-1}$ in the *T. pallidum* spectrum.

Figure 3 shows that the RR spectrum of the *T. pallidum* E48A mutant at pH 8.5 is similar to that of wild-type. Some small changes in relative band intensities (Fig. 3) could be attributed to differences in the electronic absorption properties (Fig. 1A). Closer inspection of the *T. pallidum* wild-type and E48A spectra also indicate minor frequency shifts in the 658-662/660 $cm^{-1}$ bands. These observations suggest that the ligand in the sixth position, presumably the Glu 48, appears to act as a rather weak ligand which does not sizeably perturb the vibrational structure of the active site.

**The trapped $Fe^{3+}$-peroxo species in the *T. pallidum* E48A SOR mutant.**

Figure 4 shows the resonance Raman spectrum excited at 647.1 nm of the *T. pallidum* E48A mutant at pH 8.5 treated with 6 equivalents of $H_2O_2$ then rapidly frozen (3 sec). This RR spectrum exhibits new bands at 852 and 433 $cm^{-1}$, compared with the $K_2IrCl_6$-oxidized sample (Fig. 3), which were not observed when other oxidants were used. These frequencies are consistent with the $ν_{O\text{-}O}$ and $ν_{Fe\text{-}O2}$ stretching modes, respectively, of an $Fe^{3+}$-peroxo species, as was reported for the homologous E47A mutant of SOR from *D. baarsii* [25]. Isotopically labelled $H_2{}^{18}O_2$ experiments show that these bands downshift to 802 (-50) and 413 (-20) $cm^{-1}$, respectively, in agreement with the calculated isotopic shifts of -49 and –20 $cm^{-1}$ for a Fe-$O_2$ system, respectively; this close agreement indicates that these Fe-peroxo modes are pure. $D_2O$ buffer exchange measurements were complicated by the fact that the protein was not stable in such a buffer on ice. Significant loss in metal-binding and resulting luminescence precluded the recording of good quality RR spectra. Nevertheless, such spectra (see Supplemental Data, Fig. S3) show that the 852 $cm^{-1}$ band frequency is unaffected (the 433 $cm^{-1}$ band was severely distorted by the luminescence background), suggesting no involvement of protons.

When 6 equivalents of $H_2O_2$ is rapidly mixed with wild-type SOR from *T. pallidum* also at pH 8.5, the same new 852 and 433 $cm^{-1}$ bands are also observed, however their intensities relative to the 746 $cm^{-1}$ band (a marker band for the active site in the high-spin $Fe^{3+}$ state) [25] are much weaker compared to the E48A mutant (Fig. 5). These observations demonstrate that the same high-spin side-on $Fe^{3+}$-peroxo species could be trapped for the wild-type, however with a much lower yield compared to total $Fe^{3+}$ formed than in the E48A mutant.-

A high spin $Fe^{3+}$-($\eta^2$-OO) species was also found for the WT and E47A mutant of the SOR from *D. baarsii* [25]. Interestingly, the two $Fe^{3+}$-($\eta^2$-OO) species of *D. baarsii* and *T. pallidum*, are not identical (Fig. 5). The slightly higher frequency observed for the $\nu_{O-O}$ stretching mode for *T. pallidum* (852 $cm^{-1}$) indicates that this bond is slightly stronger than it is for *D. baarsii* (850 $cm^{-1}$) [25]. In addition, the lower frequency of the $\nu_{Fe-O2}$ mode (433 $cm^{-1}$) indicates weaker $Fe-O_2$ bonding for the *T. pallidum* SOR than for that of *D. baarsii* (438 $cm^{-1}$). Thus, the $Fe^{3+}$-($\eta^2$-OO) species appears to be a weaker adduct for the SOR of *T. pallidum* than for that of *D. baarsii*. The small but significant vibrational differences observed in the *T. pallidum* $Fe^{3+}$-($\eta^2$-OO) species compared to that of *D. baarsii* mirrors the small vibrational differences observed for their respective oxidized active sites (see Fig. 3).

The formation of the $Fe^{3+}$-peroxo species in the E48A mutant of *T. pallidum* results in Raman spectral changes for the rest of the active site, both in terms of frequency shifts and relative intensities, and especially in the 250-400 $cm^{-1}$ region which corresponds to the mixed

Fe-S stretching/deformation modes. The most marked shift after peroxo formation is seen with the 304 $cm^{-1}$ band which downshifts to 297 $cm^{-1}$ for the peroxo species, suggesting a concomitant weakening of the Fe-S bond. The 357 $cm^{-1}$ band downshifts to 355 $cm^{-1}$ and the 311 and 326 $cm^{-1}$ bands slightly upshift to 313 and 328 $cm^{-1}$, respectively. The 745 $cm^{-1}$ band also upshifts very slightly to 746 $cm^{-1}$, suggesting a possible modest strengthening of the C-S bond in the cysteine residue. These observations, which are not clearly observable in the WT (Fig. 5) due to the lower yield of $Fe^{3+}$-peroxo formation, show that the peroxo ligand significantly influences the Fe-S bond in the active site, something that does not occur for the E48A mutation by itself.

The low frequency region 200-260 $cm^{-1}$ shows no significant differences from that of the E48A mutant RR spectrum oxidized with $K_2IrCl_6$, providing no evidence for significant changes in the Fe-His coordination of the active site after $Fe^{3+}$-peroxo formation. Similar observations were noted for the E47A mutant of *D. baarsii* [25].

**The SOR-Ferrocyanide Complex**

$K_3Fe(CN)_6$ is known to oxidize the reduced active site $Fe^{2+}$ centre of SOR and to form a complex with it [10,13,27,29]. This is not the case for other oxidants such as $K_2IrCl_6$ [13]. The SOR-ferrocyanide complex is described as one where one $CN^-$ group is coordinate with the oxidized SOR active site, $_{SOR}Fe^{3+}$-CN-$Fe^{2+}$-$(CN)_5$ [27,29]. The effects of this complexation are readily observable in the RR spectrum of *T. pallidum* and that of *D. baarsii*, both similar for wild-type and the E48A/E47A mutants. For simplicity we will restrict our discussion to the wild-type proteins. Upon ferrocyanide complexation, the intensity in the low frequency region is enhanced, resulting in a concomitant sharpening of the bands in the 200-260 $cm^{-1}$ (Fig. 6). This probably reflects the change in the electronic absorption properties of the active site upon ferrocyanide complexation (Fig. 1, and see Supplemental Data Fig. S2) [27], with the appearance of the intervalence broad absorption band at ca. 1000 nm that changes the resonance Raman enhancement conditions with 647.1 nm excitation.

By examining the accompanying RR changes in the Fe-S and Fe-N vibrational modes, the SOR-ferrocyanide complex serves as a sensitive probe for the active site. The effects of ferrocyanide complexation also result in more significant perturbations of the vibrational structure of the SOR active sites for *T. pallidum* and *D. baarsii* (Fig. 6). In general, many bands, including those attributable to the Fe-S(Cys) stretching modes, are seen to downshift by 1-3 $cm^{-1}$. These shifts are more pronounced for *D. baarsii* compared to *T. pallidum* and suggest a weakening of the *trans* Fe-S(Cys) bond for both SORs. These perturbations are also most notably seen for the S-C stretching mode of the cysteine ligand. For the *T. pallidum* SOR the 746 $cm^{-1}$ band downshifts to 744 $cm^{-1}$ for the SOR-ferrocyanide complex (Fig. 6). For *D. baarsii*, this band is observed at 742 $cm^{-1}$ and downshifts to 739 $cm^{-1}$ when ferrocyanide is complexed. The slightly greater downshift for *D. baarsii* (-3 $cm^{-1}$) than for *T. pallidum* (-2 $cm^{-1}$) suggests a greater perturbation or a stronger complex formation for the former compared to the latter.

Other perturbations are seen at the level of the $Fe^{3+}$-N(His) bands. For *D. baarsii*, oxidation with ferricyanide results in the upshift of the 238 $cm^{-1}$ band to 252 $cm^{-1}$. For *T. pallidum*, the analogous band at 255 $cm^{-1}$ upshifts slightly to 257 $cm^{-1}$. We also observe the downshift of another band attributable to an $Fe^{3+}$-N(His) stretching mode, 234 $cm^{-1}$ to 225 $cm^{-1}$ for *D. baarsii* and 232 $cm^{-1}$ to 230 $cm^{-1}$ for *T. pallidum*. Again, the ferrocyanide complexation effects appear more pronounced for *D. baarsii*, implying stronger complexation for this case. The X-ray crystal structure of the the *D. baarsii* SOR complexed with ferrocyanide [29] indicated that one of the $CN^-$ groups is in strong van der Waals contact with at least one of the coordinating His ligands.

## DISCUSSION

To date, three X-ray crystallographic structures are available for superoxide reductases, two for Class I SOR [9,29] and one for Class II SOR [12]. Although these crystal structures indicate that the non-heme iron active site is very similar for these SORs of different classes, they are not able to reveal subtle structural variations that might exist between one class and another. In this work, using resonance Raman spectroscopy, we have shown that the active sites of two SORs from different classes, i.e. *D. baarsii* and *T. pallidum*, exhibit significant differences, which may have important mechanistic relevance.

### Comparison of *T. pallidum* and *D. baarsii* active sites

Comparison of the RR spectra of the $K_2IrCl_6$ oxidized active sites of the *T. pallidum* and *D. baarsii* SORs indicated that these two sites are not strictly identical. These structural differences are reflected in their respective resonance Raman spectral bands as variations in both relative intensities and observed frequencies. Due to the extensive kinematic coupling and mixing of the $Fe^{3+}$-S(Cys) stretching and deformation modes in these SOR active sites [10,28], even slight changes in vibrational frequencies may significantly alter the band intensity patterns. It should be noted that with our experimental conditions, small frequency differences up to 2 $cm^{-1}$ for homologous bands on different samples are significant.

For *T. pallidum*, the cluster of bands (304, 311, 326 $cm^{-1}$) attributable to the mixed Fe-S stretching and deformation modes is observed at slightly higher frequencies than that of *D. baarsii* (299, 316, 323 $cm^{-1}$). This suggests that the Fe-S bond is stronger for the *T. pallidum* active site. Another marked difference is seen in the higher frequency band assigned to the S-$C_\beta$ stretching mode of the cysteine ligand which is seen at 746 $cm^{-1}$ for *T. pallidum* and at 742 $cm^{-1}$ for *D. baarsii*. Another interesting difference between these two SORs is seen in the region 200-260 $cm^{-1}$, which contain the contributions of $Fe^{3+}$-N(His) stretching modes [28].

The most notable is the presence of a 255 $cm^{-1}$ band in the *T. pallidum* spectrum which is seen at 238 $cm^{-1}$ for *D. baarsii*. The higher frequency seen in the *T. pallidum* spectrum indicates that at least one histidine is engaged in a significantly stronger coordination with the iron metal center, and thus a relatively shorter distance, as compared to *D. baarsii*.

To summarize, our data show that, compared to the active site of *D. baarsii*, for *T. pallidum*, i) the Fe-S bond and the cysteine S-C bond are both stronger, indicating greater electron density at these two cysteine bonds, and ii) one or two of the Fe-His bonds are stronger.

The above structural aspects might be correlated with the EPR observations [2,13,27]. With respect to the oxidized active site, *D. baarsii* exhibits a homogeneous rhombic EPR signal whereas the heterogeneous *T. pallidum* EPR spectrum exhibits species which are more axial in their iron coordination. The rhombic EPR spectrum of *D. baarsii* (Class I) might be related to its weaker $Fe^{3+}$-S(Cys) ligation and (a) weaker $Fe^{3+}$-N(His) bond(s) at the equatorial position as compared to *T. pallidum*. This could be also the case for the *D. desulfuricans* (another Class I) SOR which exhibits a rhombic EPR signal [16] while that for *P. furiosus* (Class II) is axial [13]. In fact, the *D. desulfuricans* Raman spectrum [16] is more like *D. baarsii* and the *P. furiosus* Raman spectrum [28] is more like *T. pallidum*, especially when comparing of the 742-748 $cm^{-1}$ band of the S-C cysteine stretching mode.

The structural origin of these observations is not clear yet. The available X-ray crystal structures, of differing resolutions (from 1.15 to 1.9 Å) [9,12,29], do not show the subtle differences revealed by RR spectroscopy for SOR active site in the ferric state. Even the deemed anomalously long distances of the Fe-S bond from the crystal structure of *P. furiosus* [12] might come from the incomplete Fe occupancy of the four sites in the crystal structure, and thus might be not relevant, as was discussed by Johnson and co-workers [10,13,28]. We propose that differences in the conformation of the conserved tetrapeptide sequence beginning with the active site Fe-coordinating cystein and/or subtle variations in H-bonding at the sulphur atom of the cysteine of these two SORs [10] could, in part, explain the differences observed here. For example, the additional electron density on the cysteine ligand resulting in a stronger Fe-S bond observed for the *T. pallidum* SOR active site could be reflecting less H-bonding and/or electrostatic interaction with the cysteine sulphur atom. The effects of protein environment on the thiolate axial ligand electron donation and its influence on the trans ligand properties for cytochrome P450s are well known [review 40, 41]

## SORs of different classes exhibit different stabilisation for the high-spin $Fe^{3+}$-($\eta^2$-OO) species

The reaction of the SOR from *T. pallidum*, E48A mutant or WT forms, with slight excess of $H_2O_2$ results in the formation of a high spin non-protonated side-on $Fe^{3+}$-($\eta^2$-OO) species which can be trapped upon rapid mixing and freezing. Thus, the active site of SOR from *T. pallidum* is capable of stabilizing a $Fe^{3+}$-peroxo species in a manner similar to that previously observed for *D. baarsii* WT and E47A mutant SOR [25], revealing an important general property of SORs.

When we compare the $Fe^{3+}$-peroxo species in *D. baarsii* and *T. pallidum*, there are significant differences in the observed $\nu_{O\text{-}O}$ and $\nu_{Fe\text{-}O2}$ vibrational frequencies. For *T. pallidum* the former frequency is slightly higher (852 $cm^{-1}$ compared to 850 $cm^{-1}$ for *D. baarsii*) while the latter is slightly lower (433 $cm^{-1}$ compared to 438 $cm^{-1}$). This implies a slightly stronger adduct formation for *D. baarsii*. In addition, the difference in $Fe^{3+}$-peroxo adduct strengths for *T. pallidum* and *D. baarsii* appears to correlate with the strength of the Fe-S bond in the *trans* position. For *T. pallidum*, where the Fe-S bond appears slightly stronger compared to *D. baarsii*, the $Fe^{3+}$-O bonds are weaker than those observed for *D. baarsii*.

These observations are in line with the RR data obtained for the SOR-ferrocyanide complexes. In both proteins, binding of ferrocyanide results in a frequency decrease of the Fe-S stretching modes, but to a greater extent for the *D. baarsii* enzyme. The RR data are also in agreement with the crystal structure of the complex, where binding of ferrocyanide results in the S-Fe coordination being elongated by 0.3 Å [29].

The difference in bond strengths observed for the two side-on $Fe^{3+}$-peroxo species may be rationalized in terms of electron density donation from the *trans* cysteinate ligand [30]. For a side-on bonding configuration, the in-plane π* orbital of the peroxide ligand overlaps with the iron, presumably $d_{xz}$, orbital as shown in Scheme 1. Any donation of electron density to the filled π* orbital via the iron from donation from the *trans* cysteinate ligand would be expected to weaken the $Fe-O_2$ bond. In addition, decreased $Fe-O_2$ bond strength may be expected to increase the O-O bond strength [37]. On the basis of these bond strength differences it appears that the peroxo species is more destabilized for *T. pallidum* than in *D. baarsii* and should further facilitate $Fe-O_2$ bond cleavage for this SOR, promoting better peroxide release. These RR data highlight that the different classes of SOR, which were initially classified based on the presence or absence of the N-terminal domain (and the additional Centre I), can be also now defined on the basis of different electronic distribution properties amongst the ligands of their active site, which directly affects the stability of the $Fe^{3+}$-peroxo species. This suggests that the different classes of SOR could exhibit catalytic properties which are not identical.

However, up to now, no kinetic data are available to verify that the observed weaker adduct formation for *T. pallidum* could be related to facilitating release of the $Fe^{3+}$-peroxo intermediate species to form $H_2O_2$. In fact, pulse radiolysis experiments carried out on *D. baarsii* and *T. pallidum* did not provide any information related this last step of the reaction mechanism, because of technical limitations at reaction times longer 20-30 ms. The data for *T.*

*pallidum* presented here indicate that its reaction with superoxide anion ($k_1$) is similar to that of *D. baarsii* (Table 1) and that two reaction intermediates are also formed. There is a notable difference on the rate constant for the formation of the second intermediate ($k_2$) at pH 7.6, faster for *T. pallidum*, but this step, characterized as a protonation process, was not associated with $H_2O_2$ release [24]. Other rapid kinetic techniques, like stopped-flow experiments, may provide valuable information on the release of $H_2O_2$.

At present it is not known if the side-on peroxo species is an initial SOR intermediate as seen in the pulse radiolysis experiments [21-24], given the shorter time-scales of those experiments compared to our trapping experiments with $H_2O_2$. However, the observation of a $Fe^{3+}$-($\eta^2$-OO) peroxo species in SOR reflects important properties of the active site which might correlate with an efficient release of $H_2O_2$. Recent theoretical calculations favour a linear $Fe^{3+}$-peroxo intermediate, with a low-spin state [34]. However, in all SORs studied so far, the ferric state is seen to be high-spin. For $Fe^{3+}$ complexes with aminopyridine ligands, the literature reveals that side-on $Fe^{3+}$-peroxo species have only been observed for high-spin iron model complexes so far [31]. The high-spin side-on configuration favours weaker Fe-O bonds and stronger O-O bonds compared to linear low-spin $Fe^{3+}$-($\eta^1$-OO(H)) hydroperoxo species as is revealed by the very low $\nu_{Fe\text{-}O2}$ stretching mode frequencies for side-on configurations [31]. In terms of the function and activity of SOR, an important step is the release of $H_2O_2$ from a transient $Fe^{3+}$-peroxo intermediate without cleavage of the O-O bond. To optimise such a reaction, the protein should accommodate a peroxo intermediate which has a weak Fe-O bond and a strong O-O bond. The high-spin side-on peroxo configuration meets all these requirements, better than a linear one. Indeed, linear end-on $Fe^{3+}$-OO(H) peroxo intermediates, as seen in almost all cases to be low spin $Fe^{3+}$ [31], have been observed or proposed as intermediates for catalase, cytochrome P450, NO synthase etc. These linear peroxo intermediates would result in a relatively strong Fe-O bond and a significant weakening of the O-O band, facilitating cleavage of the O-O bond to generate a highly oxidizing intermediate (e.g. (Fe(IV)=O)) required for the oxygenation activity of these latter two enzymes [36]. The low frequencies of the $Fe-O_2$ vibrations seen for high spin side-on peroxo species observed here indicate particularly weak $Fe-O_2$ bonds which would facilitate $H_2O_2$ release and stabilize the O-O bond against cleavage

*The Glu47/48 residue as sixth ligand*

According to the X-ray crystal structure of the *P. furiosus* SOR [12] and FTIR spectroscopic studies on the *D. baarsii* and *T. pallidum* SORs [14], the Glu47/48 residue appears to act as the sixth ligand of coordination to the $Fe^{3+}$ in the active site after oxidation. The resonance Raman spectra reported here using 647.1 nm excitation were not expected to show significant enhancement of any mode(s) from a possible Glu residue since there is no corresponding charge transfer band to exploit in this spectral region.

Our recent Mössbauer study [26] for *D. baarsii* indicated near 100% trapping of the side-on peroxo species at the active site of the E47A mutant, when it is formed using the same experimental conditions as reported here. Comparing the resonance Raman spectra, it appears that near 100% trapping of the $Fe^{3+}$-peroxo species in the E48A *T. pallidum* mutant was also obtained here. Thus, in the absence of the glutamic acid residue, a metastable $Fe^{3+}$-peroxo species can be readily formed and trapped by treatment with $H_2O_2$ and rapid freezing. We note that the same $Fe^{3+}$-peroxo species could also be trapped for the corresponding WT SORs but with a much lower yield. These observations again highlight the participation of the conserved Glu residue in the release of peroxide in the active site, a process which would be facilitated if the Glu residue is somehow involved, either directly or indirectly, together with the protonation and release of the peroxo species to produce $H_2O_2$.

In summary, we have observed structural differences between the active sites of *T. pallidum* and *D. baarsii* which indicate greater electron density at the cysteine ligand of *T. pallidum*. This translates as observed differences of the Fe-$O_2$ and O-O bond strengths for the trapped high-spin side-on $Fe^{3+}$-peroxo species of these two SORs which are interpreted as resulting from differences in the electron density donating properties of the *trans* cysteine ligand. The *trans* cysteine ligand effect is expected to play an important role in SOR activity with respect to superoxide reduction and $H_2O_2$ release. The high-spin side-on $Fe^{3+}$-peroxo species we have characterized here reveals that SORs in general are capable of stabilizing such species and in this light, it may be considered as a potential candidate for a subsequent intermediate in the SOR catalytic cycle.

**ACKNOWLEDGEMENTS**

CM and TAM thank Drs. V. Balland and J. Santolini for helpful discussions and Dr. A.W. Rutherford for continued support of this work. We thank Dr. Stéphane Ménage for EPR experiments and discussions, and Dr. Vincent Favaudon for assistance in pulse radiolysis experiments. We are grateful to Prof. Marc Fontecave for constant support in this work.

**SUPPLEMENTAL DATA AVAILABLE**

15 K resonance Raman spectra of *D. baarsii* SOR E47A mutant in the ferrous and ferric states treated with $H_2O_2$; UV-visible absorption spectra of oxidized wild type *T. pallidum* SOR; 15 K resonance Raman spectra of E48A *T. pallidum* SOR in $H_2O$ and $^2H_2O$ buffers and oxidized with $H_2O_2$.

**Table 1: Comparison of the observed rate constants $k_1$ and $k_2$ for the reaction of SORs from *T. pallidum* and *D. baarsii* with superoxide, determined by pulse radiolysis at pH 7.6.**

$$SOR\text{-}Fe^{2+} + O_2^{\bullet -} \xrightarrow{k_1} 1^{st}\text{ intermediate} \xrightarrow{k_2} 2^{nd}\text{ intermediate} \rightarrow SOR\text{-}Fe^{3+} + H_2O_2$$

| SOR | *T. pallidum* wild-type from Ref. [21] | *T. pallidum* E48A this work | *D. baarsii* wild-type from Ref. [20] | *D. baarsii* E47A from Ref. [20] |
|---|---|---|---|---|
| $k_1$ ($M^{-1}s^{-1}$) | $(6.0\pm0.8)x10^8$ | $(6.2\pm0.5)x10^8$ | $(1.1\pm0.3)x10^9$ | $(1.2\pm0.2)x10^9$ |
| $K_2$ ($s^{-1}$) | 4800±600 | 2080±220 | 500±50 | 440±50 |

## FIGURE LEGENDS

**Figure 1. A. UV-visible absorption spectra of the SOR from *T. pallidum* E48A mutant (200 μM).** Proteins oxidized with (a) 3 equivalents of $K_2IrCl_6$ at pH 5.0, (b) 3 equivalents of $K_2IrCl_6$ at pH 8.5, (c) 6 equivalents $H_2O_2$ at pH 8.5 (incubation time 15 sec), (d) 1.5 equivalents $K_3Fe(CN)_6$ pH 8.5. B. **pH dependence of the absorbance at 660 nm**. The titration curve fitted the equation expected from a single protonation process, $A_{660} = (A_{660max} + A_{660min} \times 10^{(pH-pKa)})/(1+10^{(pH-pKa)})$.

**Figure 2. EPR spectra of the SOR from *T. pallidum* E48A mutant (200µM).** Proteins oxidized with (a) 3 equivalents of $K_2IrCl_6$ at pH 8.5, (b) 6 equivalents $H_2O_2$ pH 8.5 (incubation time 15 sec). EPR conditions: temperature, 4.2 K; microwave power 25 mW at 9.447 GHz; modulation amplitude, 20G at 100 kHz. The Inset shows a blow up of the 500-600 Gauss magnetic field region.

**Figure 3. Low temperature (15 K) resonance Raman spectra of SORs excited at 647.1 nm.** (a) *T. pallidum* E48A mutant, (b) *T. pallidum* WT, (c) *D. baarsii* WT, all oxidized with 3 equivalents of $K_2IrCl_6$. (c) is the difference spectrum of oxidized-minus-untreated (reduced), as described in the text. SOR concentration was 3 mM in 100 mM Tris-HCl pH 8.5; laser power was 50 mW at the sample; spectral resolution was < 3 $cm^{-1}$. Raman contributions of ice were also subtracted.

**Figure 4. Low temperature (15 K) resonance Raman spectra of SOR active site from *T. pallidum* E48A mutant oxidized with 6 equivalents of $H_2O_2$, rapidly mixed and frozen within 3 sec.** (a) $H_2{}^{16}O_2$; (b) $H_2{}^{18}O_2$; (c) is the calculated difference of spectrum (b)-minus-

spectrum (a). Bands indicated by (*) are due to added, unreacted $H_2{}^{16/18}O_2$. SOR concentration was 3 mM in 100 mM Tris-HCl pH 8.5. Same experimental conditions as in Figure 3.

**Figure 5. A. Low temperature (15 K) resonance Raman spectra of the $H_2O_2$-oxidized SORs.** (a) *T. pallidum* WT (3 mM), (b) *T. pallidum* E48A mutant (3 mM), and (c) *D. baarsii* E47A mutant (5 mM), in 100 mM Tris-HCl pH 8.5, rapidly mixed with 6 equivalents of $H_2O_2$ on ice, and immediately frozen within 3 sec. (c) is the calculated oxidized-minus-reduced spectrum of *D. baarsii* as described in the text. Same experimental conditions as in Figure 3. **B. Expanded view of the 700-900 $cm^{-1}$ region**. Bands indicated by (*) are due to added, unreacted $H_2O_2$.

**Figure 6. Low temperature (15 K) resonance Raman spectra of oxidized WT SORs.** (a) *T. pallidum* oxidized with 3 equivalents $K_2IrCl_6$, (b) *T. pallidum* oxidized with 1.5 equivalents $K_3Fe(CN)_6$, (c) *D. baarsii* oxidized with 1.5 equivalents $K_3Fe(CN)_6$, (d) *D. baarsii* oxidized 3 equivalents $K_2IrCl_6$. SOR concentration was 3 mM in 100 mM Tris-HCl pH 8.5. Same experimental conditions as Figure 3.

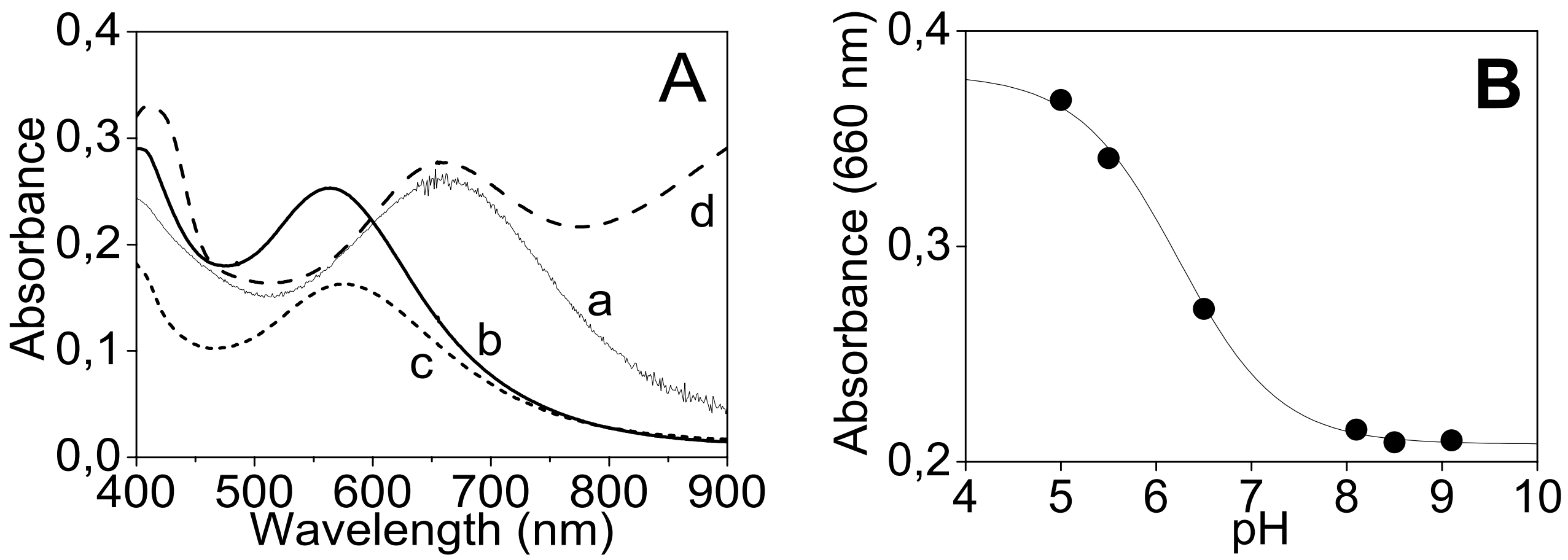

Figure 1

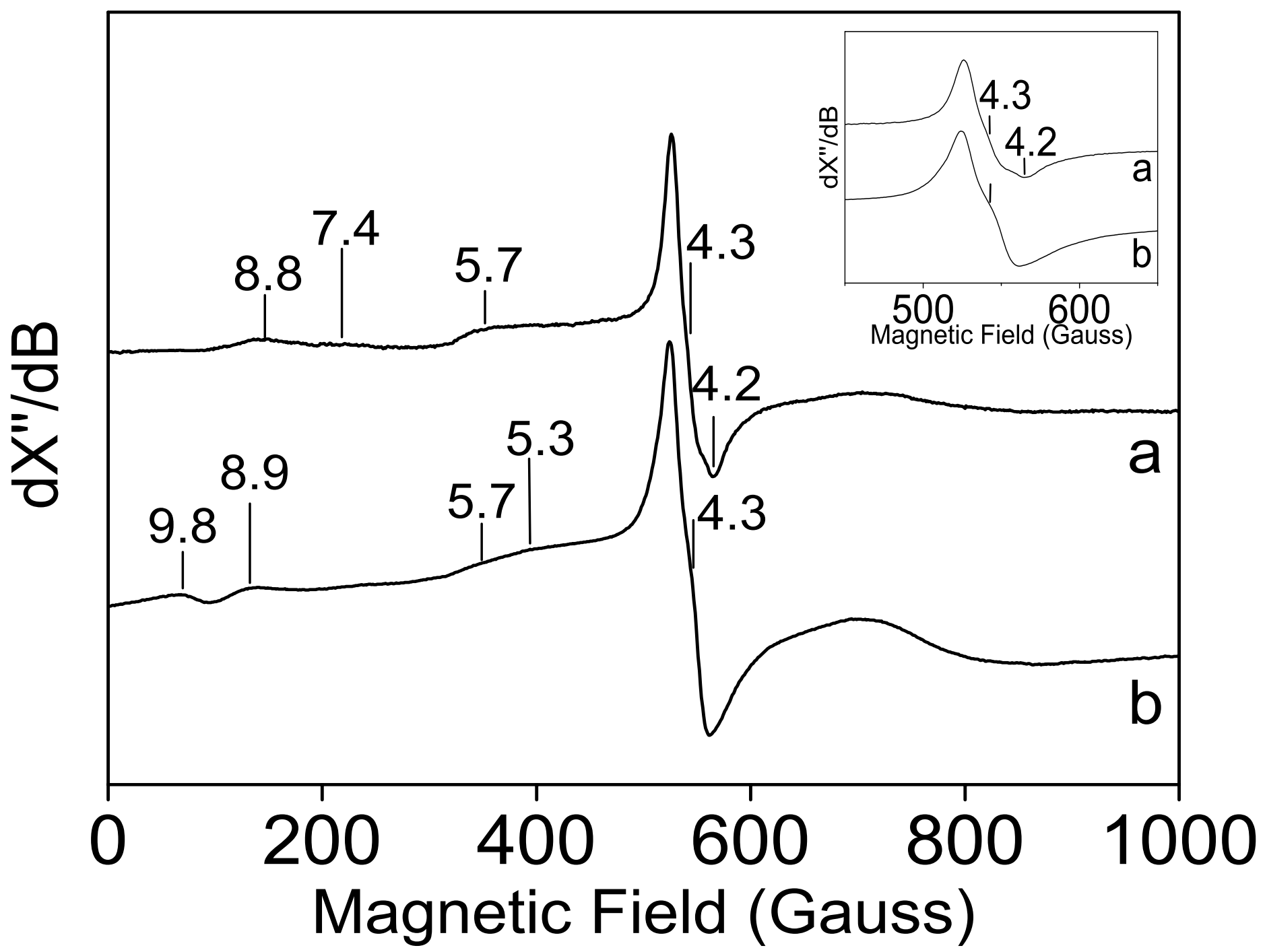

Figure 2

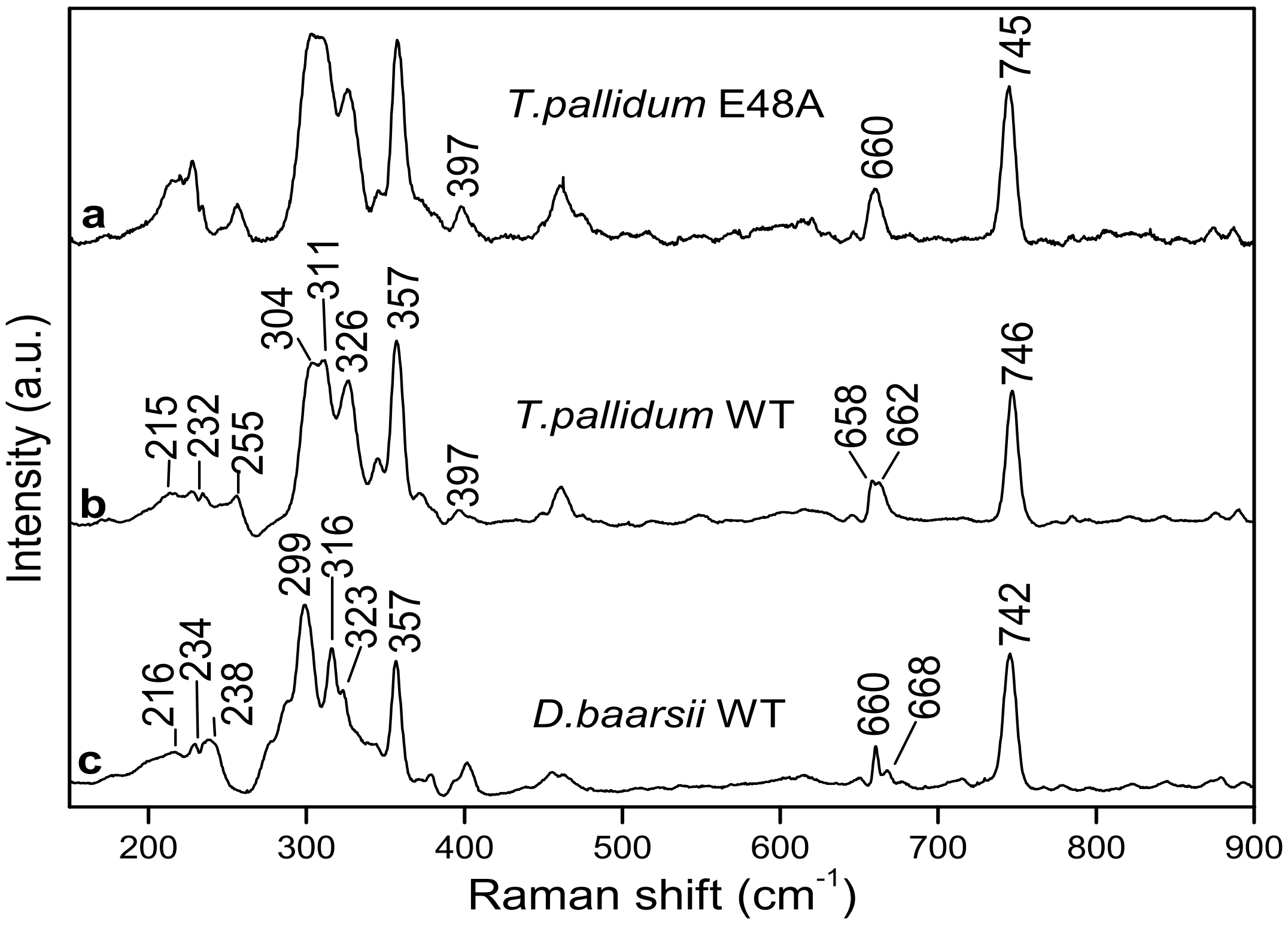

Figure 3

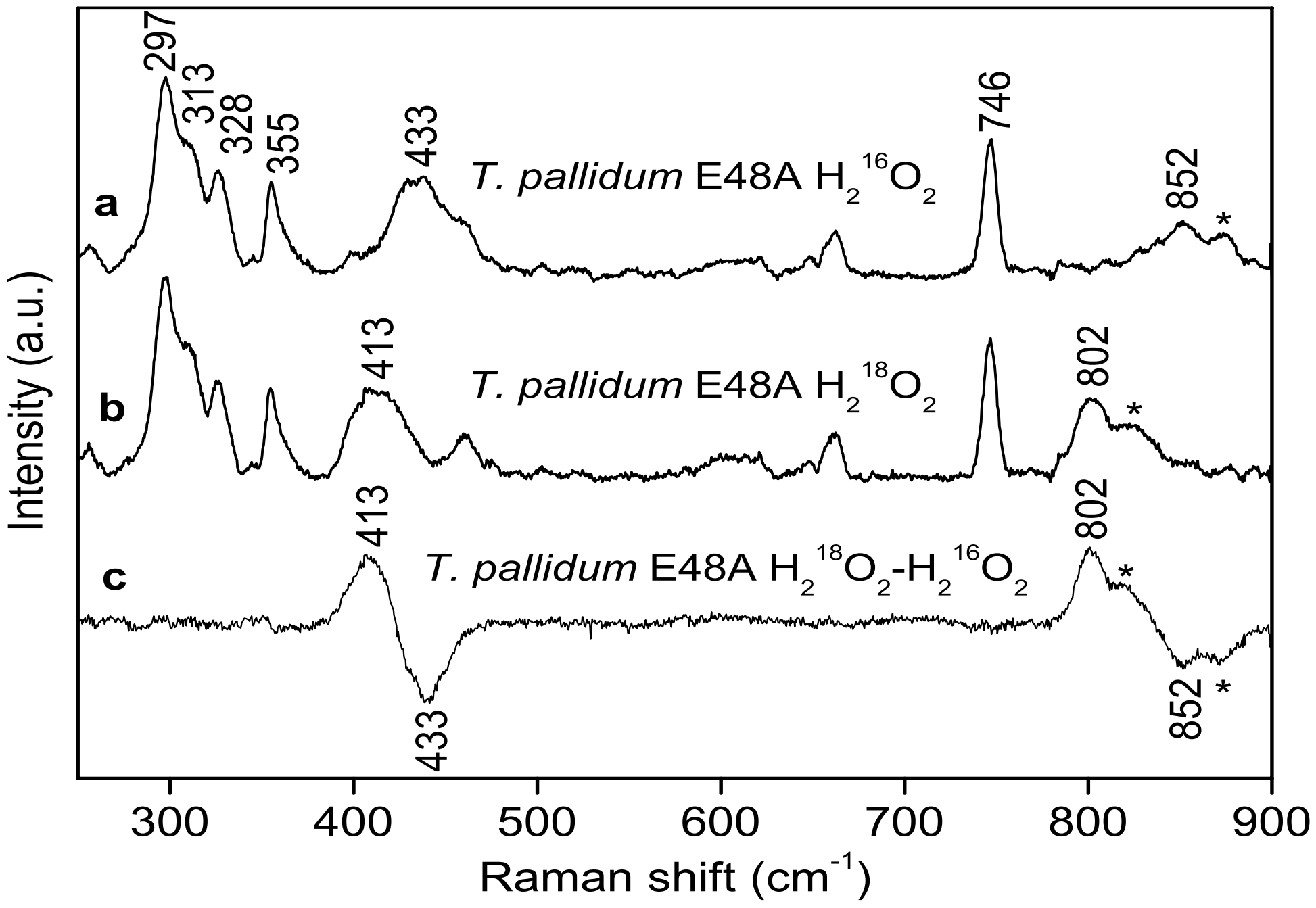

Figure 4

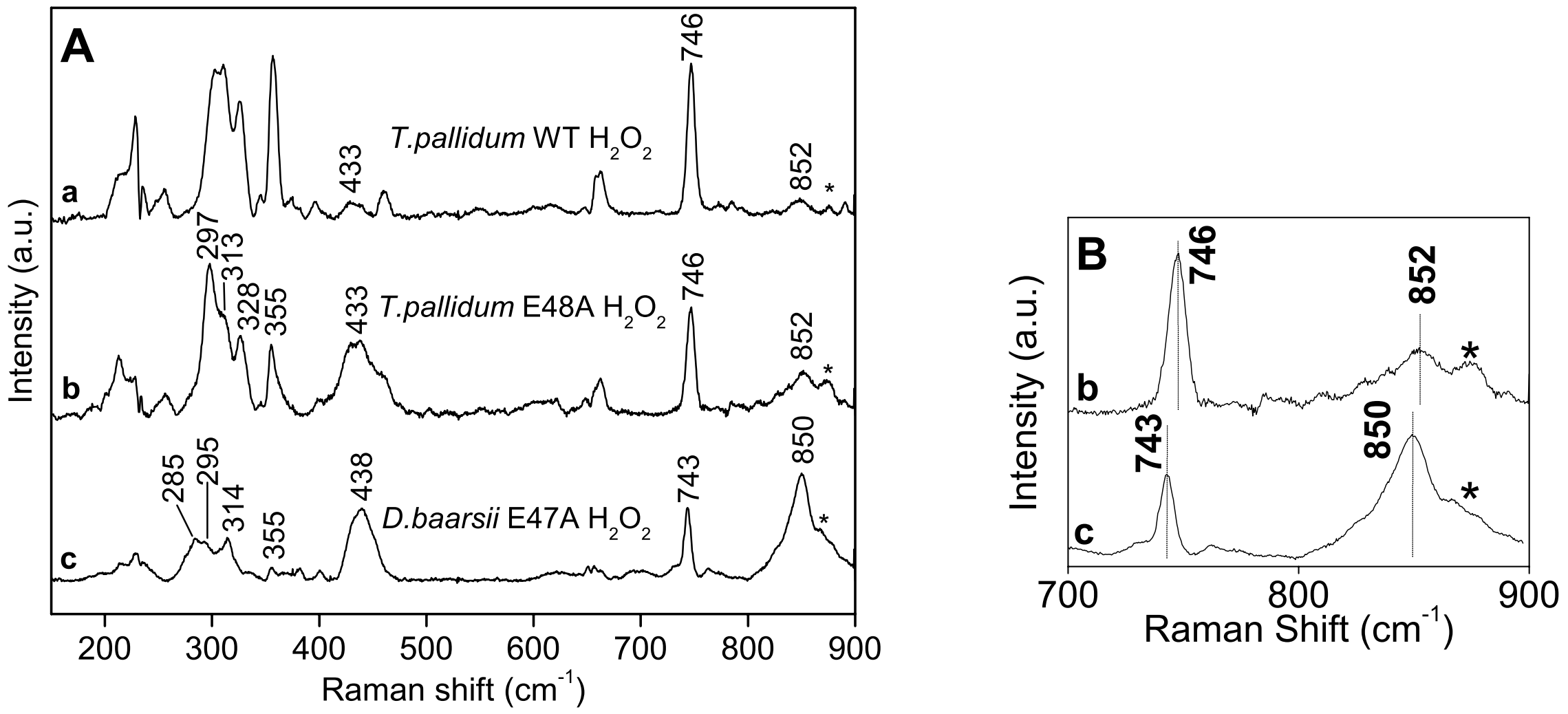

Figure 5

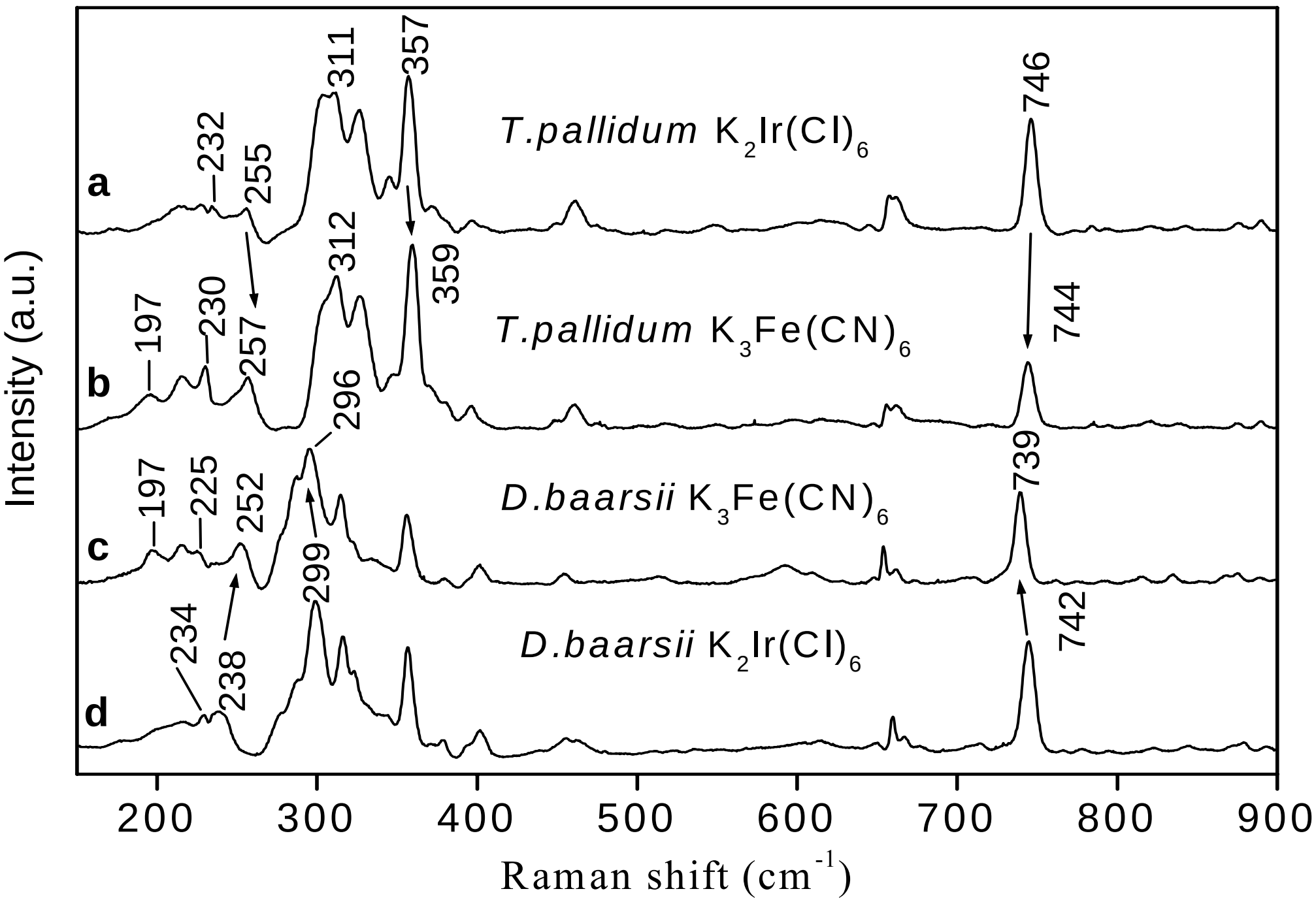

Figure 6

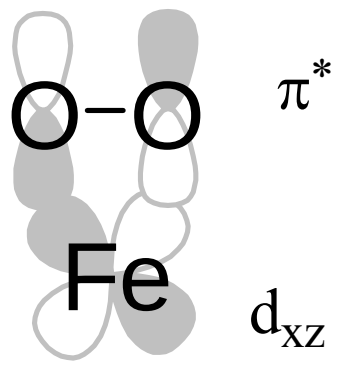

Scheme 1